\documentclass[12pt]{prop_wp} 

\usepackage[super,comma,sort&compress]{natbib} 
\usepackage{apjfonts} 
\usepackage{graphicx} 
\usepackage{amssymb,amsmath} 
\usepackage{color} 
\usepackage[table]{xcolor} 
\usepackage{float}

\usepackage{geometry}
\usepackage[utf8]{inputenc}
\usepackage{enumitem,amssymb}
\newlist{thematic}{itemize}{8}
\setlist[thematic]{label=$\square$}

\usepackage{pifont}
\newcommand{\cmark}{\ding{51}}%
\newcommand{\done}{\rlap{$\square$}{\raisebox{2pt}{\large\hspace{1pt}\cmark}}%
\hspace{-2.5pt}}

\newcommand{\axis}{{\it\small AXIS}}

\newcommand{\chandra}{\textit{Chandra}}
\newcommand{\Chandra}{\textit{Chandra}}
\newcommand{\lynx}{\textit{Lynx}}
\newcommand{\athena}{\textit{Athena}}

\newcommand{\xrism}{{\it\small XRISM}}
\newcommand{\xarm}{{\it\small XRISM}}
\newcommand{\lofar}{{\it\small LOFAR}}

\newcommand{\mwa}{{\it\small MWA}}

\renewcommand{\textfloatsep}{4mm}

\begin{document}
{\raggedright
\pagenumbering{roman}
\huge
Astro2020 Science White Paper \linebreak

Physics of cosmic plasmas from high angular resolution X-ray imaging of galaxy clusters  \linebreak
\normalsize

\noindent \textbf{Thematic Areas:} \hspace*{60pt} $\square$ Planetary Systems \hspace*{10pt} $\square$ Star and Planet Formation \hspace*{20pt}\linebreak
$\square$ Formation and Evolution of Compact Objects \hspace*{31pt} $\done$ Cosmology and Fundamental Physics \linebreak
  $\square$  Stars and Stellar Evolution \hspace*{1pt} $\square$ Resolved Stellar Populations and their Environments \hspace*{40pt} \linebreak
  $\done$    Galaxy Evolution   \hspace*{45pt} $\square$             Multi-Messenger Astronomy and Astrophysics \hspace*{65pt} \linebreak
  
\textbf{Principal Author:}

Name:	Maxim Markevitch
 \linebreak						
Institution:  NASA GSFC
 \linebreak
Email: maxim.markevitch@nasa.gov
 \linebreak
Phone:  301-286-5947
 \linebreak
 
\textbf{Co-authors:} 
Esra~Bulbul (CfA),
Eugene~Churazov (MPA, IKI),
Simona~Giacintucci (NRL), 
Ralph~Kraft (CfA),
Matthew~Kunz (Princeton),
Daisuke~Nagai (Yale),
Elke~Roediger (Hull), 
Mateusz~Ruszkowski (Michigan),
Alex~Schekochihin (Oxford),
Reinout~van~Weeren (Leiden), 
Alexey~Vikhlinin (CfA),
Stephen~A.~Walker (NASA GSFC), 
Qian~Wang (Maryland),
Norbert~Werner (ELTE, Masaryk),
Daniel~Wik (Utah), 
Irina~Zhuravleva (Chicago),
John~ZuHone (CfA)
\linebreak

\textbf{Abstract:}

Galaxy clusters are massive dark matter-dominated systems 
filled with X-ray emitting, optically thin
plasma. Their large size and relative simplicity (at least as 
astrophysical objects go) 
make them a unique laboratory to measure some of the interesting plasma 
properties that are inaccessible by other means but fundamentally
important for understanding and modeling many 
astrophysical phenomena --- from solar flares to black hole accretion to 
galaxy formation and the emergence of the cosmological Large Scale 
Structure. While every cluster astrophysicist is eagerly anticipating 
the direct gas velocity measurements from the forthcoming 
microcalorimeters onboard \xrism, \athena\ and future missions such as \lynx, 
a number of those plasma properties can best be probed 
by high-resolution X-ray {\em imaging}\/ of galaxy clusters. \chandra\
has obtained some trailblazing results, but only grazed the surface of such studies. In 
this white paper, we discuss why we need arcsecond-resolution, 
high collecting area, low relative background 
X-ray imagers (with modest spectral resolution), such as the proposed 
\axis\ and the imaging detector of \lynx.
}

\vspace*{1cm}
\noindent March 11, 2019 ~v3

\pagebreak

\twocolumn
\setcounter{page}{1}
\pagenumbering{arabic}

\noindent
{\sc Modern} astrophysics relies on computer simulations to help us 
understand complex phenomena in the
Universe, from solar flares to supernova explosions, black hole 
accretion, galaxy formation and the emergence of Large Scale Structure. As
supercomputers advance, the benefits of numeric simulations will grow.
However, for systems that include plasma, there is a fundamental limitation
--- we can't simultaneously model all the relevant linear scales from
first principles. For example, turbulence in the cosmological volume 
is driven by structure formation on the galaxy cluster scale
($10^{24}$ cm), but can cascade down to scales as small as
the ion gyroradius ($10^{8-9}$ cm), a dynamic range that is impossible
to implement in codes. To model such systems, we have to rely on 
observed plasma properties and encode them at 
the ``subgrid'' level. However, many properties that affect 
large-scale phenomena --- 
viscosity, heat conductivity, energy exchange between the particle
populations and the magnetic field --- are still unmeasured 
and their theoretical estimates uncertain by orders of magnitude 
because of the complexity of the plasma physics. Of course, apart from
being ``under the hood'' of many astrophysical systems,
plasma physics is interesting on its own.

Mircoscale phenomena in $\beta\!\sim\!1$ plasmas (where $\beta$ is 
the ratio of thermal to magnetic pressure) can be studied in situ in
our space neighborhood. Larger scales, including the transition from
``kinetic'' to ``fluid'' regime, can be probed in another natural laboratory 
that is galaxy clusters. Clusters are Megaparsec-size clouds of  
X-ray emitting, optically thin plasma (ICM), permeated by tangled magnetic fields and
ultrarelativistic particles, with typical $\beta\!>\!100$. This regime is
directly relevant for many astrophysical systems, among them SNR, 
accretion disks and the intergalactic medium.  

Several phenomena observed in clusters are sensitive to plasma physics.
Turbulence is one, and it will be characterized by the future microcalorimeters 
(\xarm\ and \athena) using Doppler shifts of the X-ray emission lines. Several
important measurements can be done using high-resolution X-ray {\em imaging}.
Shock fronts, discovered by \Chandra\ thanks to its sharp mirror, 
let us study heat conductivity, 
the electron-ion temperature equilibration and the physics of
cosmic ray acceleration\cite{Markevitch2007}. 
Another interesting plasma probe is provided by the ubiquitous, sharp contact 
discontinuities, or ``cold fronts''\cite{Markevitch2007}. While
\Chandra\ has obtained tantalizing results,
it has only scratched the surface of what can be learned from detailed imaging of these and some other cluster phenomena.

\begin{figure*}
    \centering
    \includegraphics[width=\textwidth,viewport=1 13 459 166,clip]{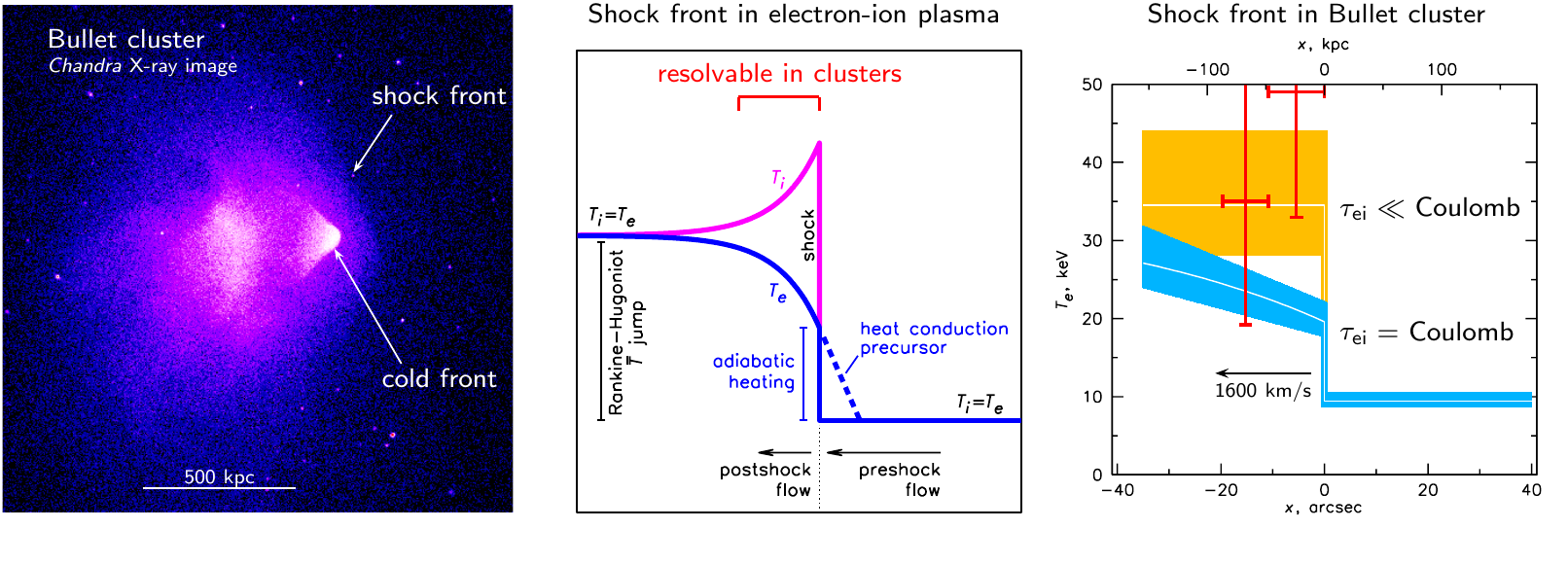}
    \caption{
    ({\em a}) X-ray image of the Bullet cluster, the textbook example of
      a bow shock. The shock is driven by a moving subcluster,
      whose front boundary is a ``cold front.'' ({\em b}) Expected 
      electron and ion temperature profiles across a shock front. 
      Temperatures are unequal immediately after the shock and then
      equalize. If electron heat conduction is not suppressed, a temperature
      precursor is also expected. ({\em c}) \Chandra{} deprojected electron
      temperature profile immediately behind the Bullet shock (crosses; 
      errors are $1\sigma$) with models for Coulomb collisional and instant equipartition\cite{Markevitch2006}. This measurement
      favors fast electron-proton equilibration, but uncertainties are large.}
    \label{fig:tei}
\end{figure*}

\section*{PLASMA EQUIPARTITION TIMES}

The common assumption that all particles in a plasma have the same local
temperature may not be true if the electron-ion equilibration
timescale is longer than heating timescales\cite{Takizawa1999,Kawazura2019}. 
This timescale is 
fundamental for such processes as accretion onto black holes and 
X-ray emission from the intergalactic medium. It can be 
directly measured using cluster shocks. 

At a low-Mach shock, ions are
dissipatively heated to a temperature $T_i$, while electrons are adiabatically compressed to a
lower $T_e$. The two species then equilibrate to the
mean post-shock temperature\cite{Zeldovich1966} (Fig.~\ref{fig:tei}). 
From the X-ray brightness and spectra, we can measure the plasma density
and $T_e$ across the shock (this requires only a modest spectral resolution). 
For the typical low sonic Mach numbers in clusters 
($M=2-3$), the mean
post-shock temperature can be accurately predicted from
the shock density jump. 
If the equilibration is via Coulomb collisions,
the region over which the electron temperature $T_e$ increases 
is tens of kpc wide --- resolvable with a \chandra-like telescope at distances of $z<2$.  
This direct test is unique to cluster shocks because of the
fortuitous combination of the linear scales and relatively low Mach numbers;
it cannot be done for the solar wind or SNR shocks.

A \Chandra{} measurement for the 
Bullet cluster shock (Fig.~\ref{fig:tei}) suggests that 
$T_e-T_i$ equilibration is quicker than 
Coulomb\cite{Markevitch2006}, although with a systematic uncertainty that arises from the assumption of 
symmetry and requires averaging over a sample of shocks.
With \chandra, this measurement is limited to only three shocks, and the results
are contradictory\cite{Markevitch2006,Russell2012,Wang2018}.
A more sensitive imager is needed to find many more 
shocks (most of them in the cluster outskirts), select a sample of 
suitable ones, and robustly determine this basic plasma property.

\begin{figure*}
    \centering
    \includegraphics[width=\textwidth]{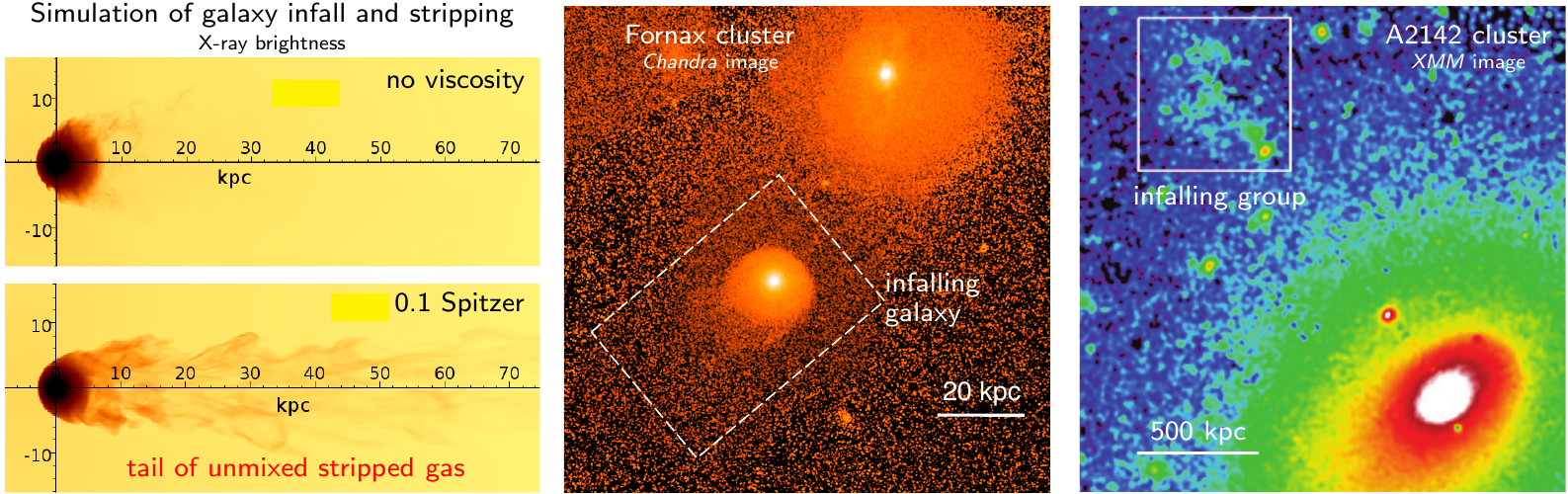}
    \caption{Plasma viscosity determines how the gas is stripped from the infalling 
    groups and galaxies. 
    {\em Left}: 
    If viscosity is not strongly suppressed, galaxies falling into clusters 
    should exhibit prominent tails of stripped gas\cite{Roediger2015}. {\em Middle, right}: An infalling galaxy (NGC1404), which appears not to have 
    such a tail\cite{Su2017}, and a much larger infalling group in the outskirts of a cluster\cite{Eckert2014}, which 
    does.}
    \label{fig:tails}
\end{figure*}

\section*{HEAT CONDUCTIVITY}

Heat conduction erases temperature gradients and competes with 
radiative cooling, and is of utmost 
importance for galaxy and cluster formation. The effective heat
conductivity in a plasma with tangled magnetic fields is unknown, 
with a large uncertainty for the component parallel to the 
field, which recent theoretical works predict to be
reduced\cite{Schekochihin2008,Kunz2014,Komarov2016,Komarov2018,RobergClark2018}. 
The existence of cold fronts in clusters confirms that
conduction across the field lines is very
low\cite{Ettori2000,Vikhlinin2001,Wang2016}, but constraints
for the average or parallel conductivity are
poor\cite{Markevitch2003,Wang2016}. 
Shock fronts are locations where the parallel component can be 
constrained, because the field lines should connect the post-shock 
and pre-shock regions (unlike for the magnetically-insulated 
cold fronts), though the field structure in the narrow shock layer 
can be chaotic. Electron-dominated conduction may result in an 
observable $T_e$ precursor (Fig.~\ref{fig:tei}).

The magnetic field can be stretched and untangled in a predictable way in the cluster sloshing
cool cores. The characteristic spiral temperature
structure that forms there\cite{ZuHone2013} can also be used to constrain parallel
conductivity. A telescope with a bigger mirror than
\chandra's could look for temperature precursors in shocks 
and obtain detailed maps of temperature gradients along the field filaments in many 
cluster cores to measure the conductivity.

\section*{VISCOSITY}

Plasma viscosity is a fundamental quantity that governs damping of turbulence and sound
waves, suppression of hydrodynamic instabilities and mixing of different gas phases, and 
thus relevant to such important processes as heating the gas, spreading metals
ejected from galaxies, and amplification of magnetic fields. At present it is largely unknown. Isotropic viscosity can be determined from the dissipation scale of the 
power spectrum of turbulence. \xrism\ and \athena\ will pursue that via the velocity measurements in the ICM, though it is
unclear if the dissipation scale will be reachable\cite{ZuHone2016}. The turbulence spectrum can also be constrained by observing the gas density fluctuations\cite{Schuecker2004,Zhuravleva2015}. However, the plasma viscosity should be anisotropic and may affect turbulence and other phenomena differently. It is thus useful to approach it from several angles. Two subtle phenomena in galaxy cluster images can help us probe the viscosity through its effect on gasdynamic instabilities.

\subsubsection*{~~Galaxy stripping tails.}

Figure~\ref{fig:tails}{\em a}\/ shows
a striking difference in the simulated X-ray appearance of the tail of the
cool stripped gas behind a galaxy as it flies through the 
ICM\cite{Roediger2015}. In an inviscid plasma, the gas
promptly mixes with the ambient ICM, but a modest viscosity suppresses the
mixing and makes the long tail visible. Deep \Chandra{} images of such infalling galaxies 
NGC1404 (Fig.\ \ref{fig:tails}{\em b}) and M89 favor efficient mixing and a reduced 
viscosity\cite{Su2017,Kraft2017}. Other infalling groups in the cluster periphery
do exhibit unmixed tails (e.g., Fig.\ \ref{fig:tails}{\em c}). This points to a 
possibility of a systematic study to constrain effective viscosity --- and directly 
observe its effect on gas mixing --- in various ICM regimes. However, a 
more sensitive instrument with lower background is required to study these subtle, 
low-contrast extended features, most of which will be found in the 
low-brightness cluster outskirts.

\begin{figure*}
    \centering
    \includegraphics[width=\textwidth]{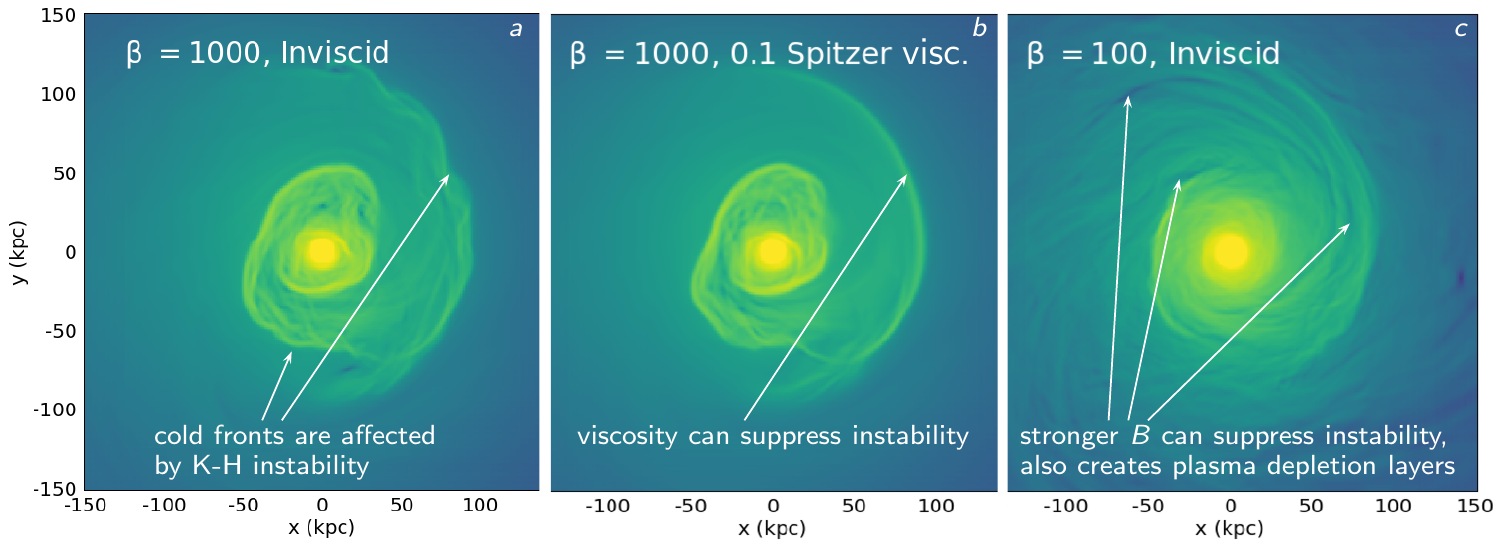}
    \caption{MHD simulation of a sloshing cluster core with viscosity (isotropic) 
    and magnetic field\cite{Bellomi2019}. X-ray brightness gradients are shown.
    Initial $\beta$ values are given; sloshing amplifies the magnetic field and produces lower 
    $\beta$, which result in plasma depletion regions. 
    The appearance of cold fronts can be used to constrain the effective plasma 
    viscosity and magnetic field strength.}
    \label{fig:cf_visc}
\end{figure*}

\subsubsection*{~~Instabilities in cold fronts.}

Cold fronts --- contact discontinuities in the ICM that separate regions of
different density and temperature in pressure 
equilibrium\cite{Markevitch2007} --- are
ubiquitous in merging subclusters, where they are seen as sharp X-ray brightness 
edges (e.g., the 
``bullet'' boundary in the Bullet  cluster, Fig.\ \ref{fig:tei}{\em a}). 
They are also found in most cool cores, where 
they emerge as the dense gas of the core ``sloshes'' in
the cluster gravitational well\cite{Ascasibar2006}. Sloshing produces velocity
shear across the cold front, which should generate Kelvin-Helmholtz instabilities 
(Fig.\ \ref{fig:cf_visc}{\em a}). If the ICM is viscous, K-H 
instabilities are suppressed\cite{Churazov2004,Roediger2013b,Zuhone2015} 
(Fig.\ \ref{fig:cf_visc}{\em b}). \Chandra\
has discovered K-H instabilities in a few cold fronts and placed an 
upper limit on the effective isotropic viscosity of $\sim0.1$ 
Spitzer\cite{Roediger2013a,Ichinohe2017,Wang2018b} 
(or, equivalently in this context, a full 
Braginskii anisotropic viscosity\cite{Zuhone2015}). 
To constrain the viscosity from below requires finding 
instabilities for a range of 
density contrasts. These subtle wiggles can be seen only
with high resolution and lots of photons, and a systematic study 
requires a larger-area telescope.


\section*{PLASMA DEPLETION LAYERS}

The velocity shear at cold fronts (and elsewhere in the cluster) should stretch and 
amplify the magnetic fields, forming magnetic layers parallel to the front. 
Such layers can suppress the instabilities even without the 
viscosity\cite{Vikhlinin2001a}, although a certain 
initial field strength is required (compare Figs.\ \ref{fig:cf_visc}{\em a,c}). 
A distinguishing feature between these two suppression
mechanisms is seen in Fig.\ \ref{fig:cf_visc}{\em c}.  Wherever the 
field is amplified, thermal plasma is squeezed out, forming 
plasma depletion layers (PDL, like the ones in the solar wind around 
planets\cite{Oieroset2004}) that can become visible in the X-ray
image\cite{Markevitch2007}. 

\begin{figure*}
    \centering
    \includegraphics[width=\textwidth]{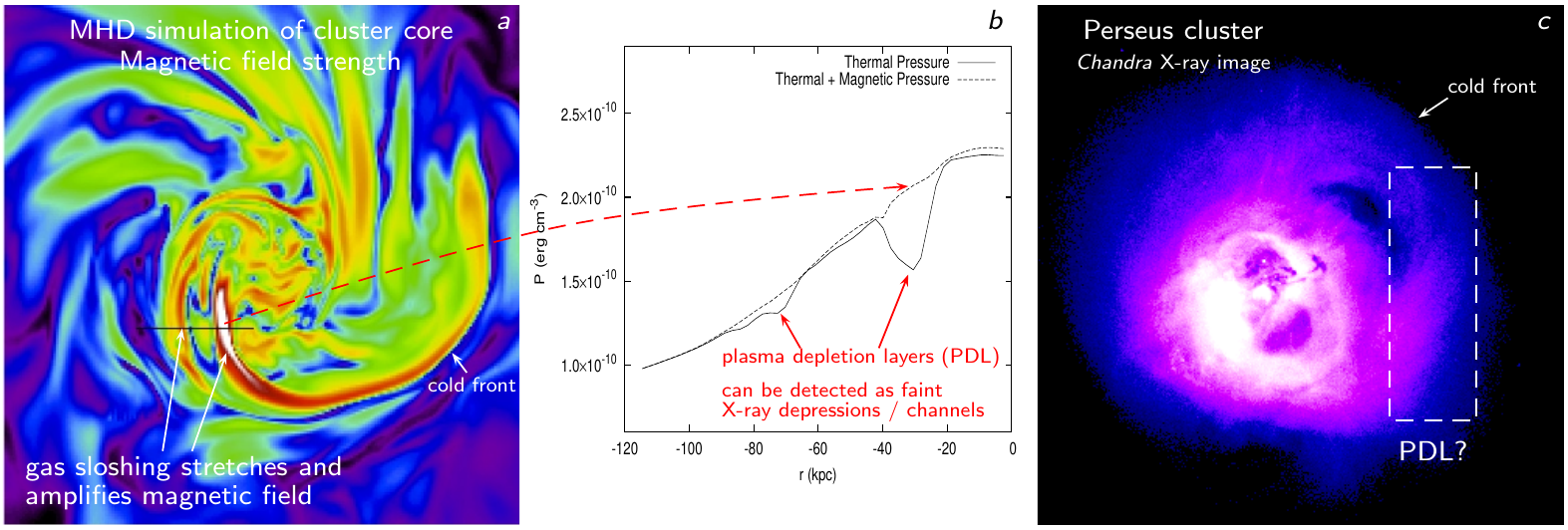}
    \caption{Plasma depletion layers in a cluster core. ({\em a}) MHD simulation of a sloshing core\cite{ZuHone2011}; color shows the field strength. As the gas swirls in the core, it forms filaments of stretched and amplified field. ({\em b}) Pressure profiles across two $\beta\sim10$ filaments, extracted along the line in panel {\em a}. While total pressure is monotonic, thermal pressure shows dips (both the density and the temperature dip). ({\em c}) Possible observation of such ``feathery'' structure in the Perseus core\cite{Ichinohe2019}. Subtle X-ray ``channels,'' possibly of similar origin, have also been seen by \chandra\ in A520\cite{Wang2016} and A2142\cite{Wang2018b}.}
    \label{fig:pdl}
\end{figure*}

In Fig.\ \ref{fig:pdl}, we show how PDL can form in a 
sloshing core. \Chandra\ has reported hints of this new phenomenon --- 
low-contrast ``channels'' in A520 and A2142\cite{Wang2016,Wang2018b} and 
``feathery'' structures in Virgo and Perseus\cite{Werner2016,Ichinohe2019}, 
Fig.\ \ref{fig:pdl}c). Apart from disentangling the effects of viscosity and 
magnetic fields on cold front stability, observing PDL in clusters would have a 
more general significance --- it allows us literally to see the structure 
of the intracluster magnetic field. Combined with the radio
images, this can map the distribution of cosmic ray electrons in the ICM.
Observing these subtle image features requires many more photons than \Chandra\ 
can collect for most clusters. A future imager with a much bigger mirror can 
give us this novel tool for cluster plasma studies.

\section*{COSMIC RAY ACCELERATION}

Across the universe, shocks accelerate particles to very high energies
via the first-order Fermi mechanism\cite{Blandford1987}. Microscopic 
details of this fundamental process remain poorly
known for astrophysical plasmas, and particle-in-cell simulations are still
far from covering realistic plasma parameters.

Many galaxy clusters exhibit striking ``radio relics'' in their 
outskirts\cite{vanWeeren2010}. These Mpc-long, arc-like 
structures are synchrotron signatures of
ultrarelativistic ($\gamma \sim 10^4$) electrons. Some relics, as well as sharp 
edges of giant radio halos, coincide with ICM shocks\cite{Giacintucci2008,Shimwell2015,Wang2018}, 
suggesting that shocks have something to do with those electrons. However, the shock 
Mach numbers are low ($M=1.5-3$) and it is unclear how they reach the acceleration
efficiency needed to produce the relics\cite{Macario2011,Brunetti2014}. 
Other puzzles include similar-Mach shocks that produce very different radio 
features\cite{Shimwell2015} and a relic for which the shock is ruled out\cite{Markevitch2019}. 
Particle acceleration in the ICM appears more complex than a classical Fermi picture. Proposed
solutions involve re-acceleration of aged relativistic particles\cite{Brunetti2014} 
as well as modifications to the Fermi mechanism in a magnetized plasma. To gain insight 
into these universal processes, we need a systematic comparison of shocks in 
the X-ray and radio. However, most radio relics are found far in the cluster 
outskirts, where the X-ray emission is too dim for \Chandra. A low-background, high-area, high-resolution X-ray imager is needed to discover and study shocks there.

\section*{FINDING MOST POWERFUL AGN OUTBURSTS}

AGN that reside in many cluster cores eject copious amounts of energy 
into the ICM, preventing runaway radiative cooling of the gas at the 
cluster centers\cite{Mcnamara2007}. 
They inflate X-ray cavities in the ICM; radio observations show 
that these cavities are filled with relativistic plasma. A recent discovery of 
a giant ghost bubble outside the core in Ophiuchus\cite{Giacintucci2019} 
suggest that the AGN effects may extend far beyond the cluster cool cores, and that AGN 
can produce far more powerful outbursts than we infer from the energetics 
of the cavities in the cluster cores\cite{Mcnamara2005}. 
If this phenomenon is widespread, as hinted at by recent 
low-freqency radio surveys by \lofar\ and \mwa, 
clusters can be affected more strongly by the AGN feedback than previously thought. 
Forensic evidence for that can be provided by large, low-contrast ghost cavities outside 
cluster cores\cite{Sanders2009}. Their detection requires a low-background, high-area X-ray imager.

\section*{WHAT KIND OF INSTRUMENT WE NEED}

All the above studies require a
much greater collecting area and much lower background than the current X-ray 
instruments can provide. 
Critically, they also require high angular resolution --- at least \chandra-like --- both 
to resolve the sharp spatial features and to remove the faint point sources of the Cosmic 
X-ray Background that dominate the flux in the cluster outskirts, 
where most of those features 
will be found. \axis, a proposed Probe, and the imaging detector 
of \lynx, a proposed Flagship, 
will have the requisite resolution and photon-collecting capabilities. They will also enable 
unsurpassed low-background imaging for $E>1$ keV (where the soft diffuse Galactic 
background becomes insignificant), as shown in the accompanying 
white paper\cite{Walker2019}.

\clearpage
\section*{REFERENCES}
\small
\parskip=0mm
\vspace*{-4mm}

\bibliographystyle{naturemag}
\bibliography{references}

\subsection*{Acknowledgments}
SG acknowledges {\em 6.1 Base funding for Basic research in radio astronomy}\/ at the Naval Research Laboratory.

\end{document}